\documentclass[english,structabstract]{aa}
\usepackage{mathptmx}
\usepackage[T1]{fontenc}
\usepackage{graphicx}

\makeatletter
\bibpunct{(}{)}{;}{a}{}{,} 
\usepackage{hyperref}
\usepackage{longtable}
\usepackage{lscape}
\newcommand{\teff}{T_{\mathrm{eff}}}
\newcommand{\logg}{\log g}
\newcommand{\feh}{\left[\mathrm{Fe}/\mathrm{H}\right]}

\newcommand{\nsad}{\vec{\nabla}_{\mathrm{sad}}}

\newcommand{\hav}{\left\langle \mathrm{3D}\right\rangle}

\newcommand{\amlt}{\alpha_{\mathrm{MLT}}}

\newcommand{\ttau}{T(\tau)}

\usepackage{txfonts}
\titlerunning{Scaling of the asymptotic entropy jump in the superadiabatic layers of stellar atmospheres}
\authorrunning{Z. Magic}

\@ifundefined{showcaptionsetup}{}{%
 \PassOptionsToPackage{caption=false}{subfig}}
\usepackage{subfig}
\makeatother

\usepackage{babel}
\begin{document}

\title{Scaling of the asymptotic entropy jump in the superadiabatic layers
of stellar atmospheres}

\author{Z. Magic\inst{1,2}}

\institute{Niels Bohr Institute, University of Copenhagen, Juliane Maries Vej
30, DK--2100 Copenhagen, Denmark \and  Centre for Star and Planet
Formation, Natural History Museum of Denmark, {\O}ster Voldgade 5-7,
DK--1350 Copenhagen, Denmark\\
\email{magic@nbi.dk}}

\date{Received ...; Accepted...}

\abstract{Stellar structure calculations are able to predict precisely the
properties of stars during their evolution. However, convection is
still modelled by the mixing length theory; therefore, the upper boundary
conditions near the optical surface do not agree with asteroseismic
observations.}{We want to improve how the outer boundary conditions
are determined in stellar structure calculations.}{We study realistic
3D stellar atmosphere models to find alternative properties.}{We
find that the asymptotic entropy run of the superadiabatic convective
surface layers exhibit a distinct universal stratification when normalized
by the entropy minimum and jump. }{The normalized entropy can be
represented by a 5th order polynomial very accurately, and a 3rd order
polynomial also yields accurate coefficients. This generic entropy
stratification or the solar stratification, when scaled by the entropy
jump and minimum, can be used to improve the modelling of superadiabatic
surface layers in stellar structure calculations. Furthermore, this
finding indicates that surface convection operates in the same way
for all cool stars, but requires further scrutiny in order to improve
our understanding of stellar atmospheres.}


\maketitle

\section{Introduction\label{sec:Introduction}}

In stellar evolution calculations, convection in the superadiabatic
region (SAR) is commonly treated with the mixing length theory \citep[MLT; see][]{1958ZA.....46..108B}.
However, as a result of the non-local and non-linear nature of surface
convection, MLT cannot correctly model the complex SAR. The bulk of
the convection zone is nearly adiabatic; i.e. compared to the adiabatic
entropy value, $s_{\mathrm{ad}}$, entropy fluctuations, $\delta s$,
are small. At the optical surface ($\tau_{\mathrm{Ross}}=1$) the
optical mean free path for a photon becomes very large, leading to
radiative losses and consequently to large-amplitude fluctuations,
essentially driving convection \citep[see][]{1998ApJ...499..914S,2009LRSP....6....2N}.
The radiative cooling at the surface leads to an entropy minimum,
$s_{\mathrm{min}}$, and determines the upper boundary of the photospheric
transition region. The resulting entropy-deficient plasma is buoyantly
accelerated downwards and subsequent mixing of the downdraft with
the stable background will rapidly diminish the entropy fluctuations
within a few pressure scale heights. Ultimately, this creates an asymptotic
entropy stratification in the SAR.

Precise stellar evolutionary calculations are important for determining
the age of stellar clusters, extragalactic population synthesis, and
the characterization of exoplanet hosts. However, the SAR is a major
source of uncertainty in stellar structure theory. This is reflected
in asteroseismology, where so-called near-surface effects have to
be corrected for, when stellar structure models are compared with
observed p-mode oscillations frequencies \citep[see][]{2008ApJ...683L.175K,2014A&A...568A.123B}.
Efforts are currently being made to implement results from 3D radiative
hydrodynamic (RHD) simulations to improve stellar structure calculations,
which is expected to enhance the accuracy of effective temperature,
$\teff$, and radius predictions \citep{2015A&A...573A..89M,2014MNRAS.442..805T,2014MNRAS.445.4366T,2015A&A...577A..60S}.
To improve MLT models, can be specified the entropy minimum and the
entropy jump, which can be done by considering calibrations of $\ttau$
relations and a variable mixing length, $\amlt$, from 3D simulations.
Nonetheless, even with this approach, MLT would still not properly
account for the true structure of the SAR. \citet{1999A&A...351..689R}
has already shown that, by appending depth dependent $\hav$ stratification
of a solar model directly onto a 1D structure, p-mode oscillation
frequency calculations can be improved considerably. Therefore, it
would be desirable to do the same for the computation of stellar structures,
even if implementing and interpolating $\hav$ stratifications onto
1D models is highly non-trivial. 

In the present work, we report on our findings regarding a universal
asymptotic stratification of the entropy jump in normalized $\hav$
entropy stratifications from the \textsc{Stagger-grid}, a grid of
3D RHD stellar atmosphere models \citep[see][]{2013A&A...557A..26M}.
These generic depth dependent entropy stratifications are easily scaled,
and can therefore potentially improve 1D stellar structure calculations.
In addition, such scaling relations are also paramount for the theoretical
understanding of surface convection.

\section{Asymptotic entropy stratification\label{sec:Entropy}}

\begin{figure*}
\subfloat[\label{fig:entropy}]{\includegraphics[width=88mm]{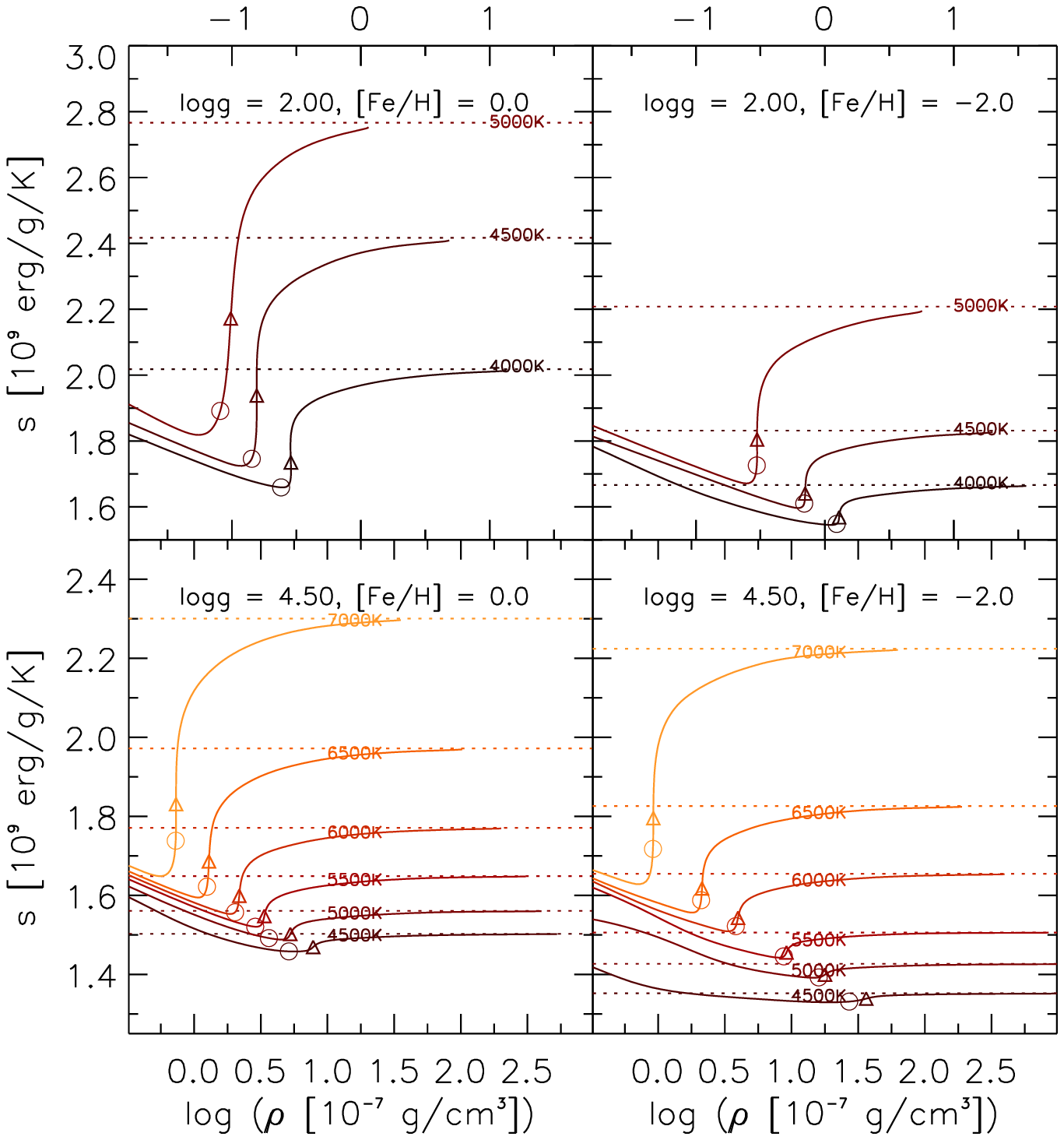}

}\subfloat[\label{fig:entropy_normalized}]{\includegraphics[width=88mm]{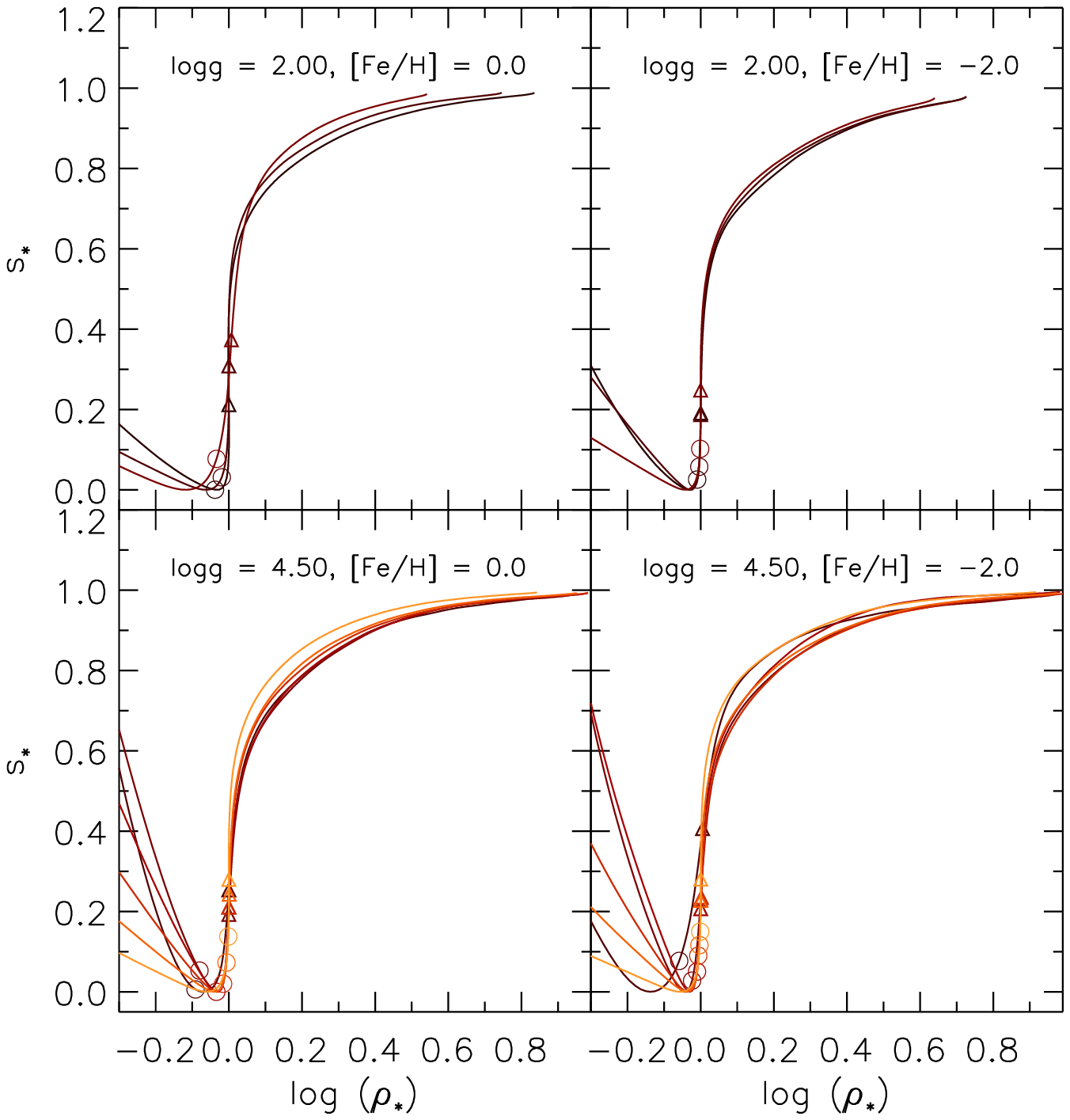}

}

\caption{\label{fig:Entropy-vs.-density}Entropy vs. density shown for different
stellar parameters \emph{without} and \emph{with} normalization (figure
(a) and (b), respectively). Each panel shows models with the same
surface gravity, $\logg$, and metallicity, $\feh$, but different
effective temperature, $\teff$ (orange/brown lines). Furthermore,
the location of the optical surface, $\tau_{\mathrm{Ross}}=1$, and
maximum in the entropy gradient, $ds|_{\mathrm{max}}$, are indicated
(circle and triangle, respectively). In the left figure, the adiabatic
entropy, $s_{\mathrm{ad}}$, values are also shown (horizontal dotted
lines). We note the differences in the axes between the top and bottom
panels in figure (a).}
\end{figure*}
The average (specific) entropy\footnote{We computed the specific (thermodynamic) entropy by integrating the
first law of thermodynamics $ds=(d\varepsilon-p_{\mathrm{th}}/\rho\,d\ln\rho)/T$.
Then we determined the spatial and temporal averages at constant geometrical
height, which are the only averages that preserve the hydrostatic
equilibrium \citep[see][for more details]{2013A&A...560A...8M}. } plotted against the density shows a very steep drop towards the optical
surface from the interior with decreasing density (Fig. \ref{fig:entropy}).
Each entropy stratification is characterized by the entropy minimum,
$s_{\mathrm{min}}$, and an asymptotic increase towards the adiabatic
entropy of the deeper convection zone $s_{\mathrm{ad}}$ with increasing
density. For higher effective temperature, $\teff$; lower surface
gravity, $\logg$; and lower metallicity, $\feh$, the $\hav$ stratifications\footnote{In the following, we label spatially and temporally averaged 3D models
over layers of constant geometrical depth with $\hav$.} tend towards lower entropy and density. Concomitantly, the entropy
jump decreases, but the asymptotic structure looks similar despite
the different depth scales. In the metal-poor case ($\feh=-2$), it
can be seen in Fig. \ref{fig:entropy} that, at high $\teff$, the
entropy jump and minimum are similar to the solar metallicity case,
while at lower $\teff$, both $\Delta s$ and $s_{\mathrm{min}}$,
are much smaller relative to the solar metallicity case. This is due
to a lack of electrons at low $\feh$ and $\teff$ that are required
for the dominant $\mathrm{H}^{-}$-opacity. At higher $\teff$, ionization
of hydrogen results in more free electrons \citep[see][]{2013A&A...557A..26M}.

We consider the following normalizations: Shifting the entropy by
its minimum and normalizing it by the entropy jump, $\Delta s=s_{\mathrm{ad}}-s_{\mathrm{min}}$,
i.e.
\begin{eqnarray}
s_{*} & = & \frac{s-s_{\mathrm{min}}}{\Delta s}.\label{eq:entropy_norm}
\end{eqnarray}
Next, normalizing the density by its value at the peak of the entropy
gradient, $ds|_{\mathrm{max}}$, and then taking the square root:
\begin{eqnarray}
\rho_{*} & = & \sqrt{\frac{\rho}{\rho\left(ds|_{\mathrm{max}}\right)}}.\label{eq:density_norm}
\end{eqnarray}
The results of the normalization are shown in Fig. \ref{fig:entropy_normalized}.
The entropy stratifications are now almost indistinguishable from
one another and can basically be scaled from one to another. The simplicity
of the outcome is remarkable, particularly when considering the complexity
of the 3D RHD simulations from where the normalized entropy jump was
derived. At the optical surface, i.e. where $\left\langle \tau_{\mathrm{Ross}}\right\rangle =1$,
the opacity and density decreases significantly, which leads to radiative
cooling in the thin photosphere\footnote{In the Sun the photosphere measures 500 km, while the convection zone
is 200 Mm deep.}, and generates the entropy fluctuations. The fluctuations are partly
advected by the horizontal deflection of velocity field, but primarily
accelerated downwards by buoyancy. The downflows shear and mix (Kelvin-Helmholtz
instability) with the layers below, which rapidly reduces the entropy
fluctuations. This causes the asymptotic convergence of the entropy
stratification. Furthermore, the process furthermore appears to be
universal. The surface gravity sets the geometrical depth scale in
the hydrostatic equilibrium, while the effective temperature, which
is given by the radiative flux, sets the height of the entropy jump.

We have also marked the locations of the optical surface and the peak
in entropy gradient in Fig. \ref{fig:Entropy-vs.-density}. One can
see that the entropy minimum, which marks the upper boundary of the
convection zone, lies well above the optical surface towards higher
effective temperature. This is due to the vigorous velocities and
overshooting, which renders the use of $\ttau$ relations, often fixed
to $\tau=2/3$ in stellar structure computations, questionable. Also,
the $\ttau$ relations are valid only for specific conditions. The
metal-poor cool dwarf (right bottom panel in Fig. \ref{fig:entropy_normalized}),
shows larger deviations from the universal entropy stratification
most likely because this model exhibits a more adiabatic stratification,
which should be considered as a limiting case for the universal entropy
scaling.

\section{Radiative cooling and convection\label{sec:Radiative-cooling}}

\begin{figure}
\includegraphics[width=88mm]{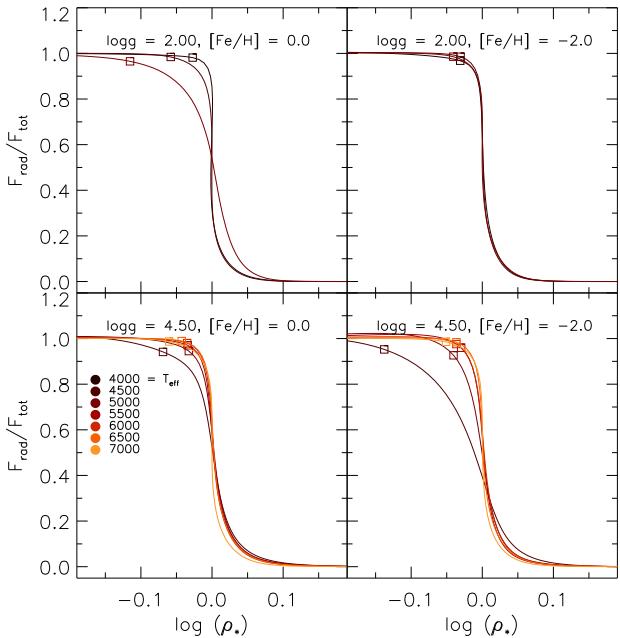}\caption{\label{fig:frad}Normalized radiative flux $F_{\mathrm{rad}}^{*}$
vs. normalized density $\rho_{*}$ for different stellar parameters.
The location of the entropy minimum is indicated (square).}
\end{figure}
Since the radiative cooling term, $q_{\mathrm{rad}}$, is responsible
for creating the entropy jump, it is worthwhile to study the radiative
flux for any additional scaling properties. When comparing the radiative
flux normalized to the total flux, $F_{\mathrm{rad}}^{*}=F_{\mathrm{rad}}/F_{\mathrm{tot}}$,
with $\rho_{*}$, then different stellar parameters exhibit the same
behaviour (Fig. \ref{fig:frad}). We find the peak of the entropy
gradient coincides with the peak of the radiative cooling gradient. 

In order to make the relation between the entropy and the radiative
flux clearer, we consider the entropy conservation equation 
\begin{eqnarray}
\partial_{t}s+\vec{u}\cdot\vec{\nabla}s & = & -\frac{1}{T}\vec{\nabla}\cdot\vec{F}_{\mathrm{rad}}+q_{\mathrm{visc}},
\end{eqnarray}
which states that the radiative losses are a result of the entropy
advection. The time derivative of the entropy becomes zero, while
the viscous dissipations are negligible. Then, the spatial and temporal
average, denoted by $\overline{\left\langle \dots\right\rangle }$,
gives
\begin{eqnarray}
2\overline{\left\langle u_{x}\partial_{x}s\right\rangle }+\overline{\left\langle u_{z}\partial_{z}s\right\rangle } & = & \overline{\left\langle q_{\mathrm{rad}}\right\rangle }/\overline{\left\langle T\right\rangle },\label{eq:entropy_advection_mean}
\end{eqnarray}
which relates the radiative cooling in the photosphere to the total
entropy advection. The horizontal entropy advection is symmetric,
i.e. $\overline{u_{x}\partial_{x}s}=\overline{u_{y}\partial_{y}s}$;
therefore, the twofold of x-direction gives the horizontal entropy
advection. The RHS of Eq. \ref{eq:entropy_advection_mean} is the
derivative of the vertical radiative flux, which determines the effective
temperature, while the LHS of Eq. \ref{eq:entropy_advection_mean}
contains the velocity, the entropy and geometrical depth scale. This
illustrates the systematic variations of these values with stellar
parameters within the \textsc{Stagger-grid}, which we reported in
\citet{2013A&A...557A..26M}.

\begin{figure}
\includegraphics[width=88mm]{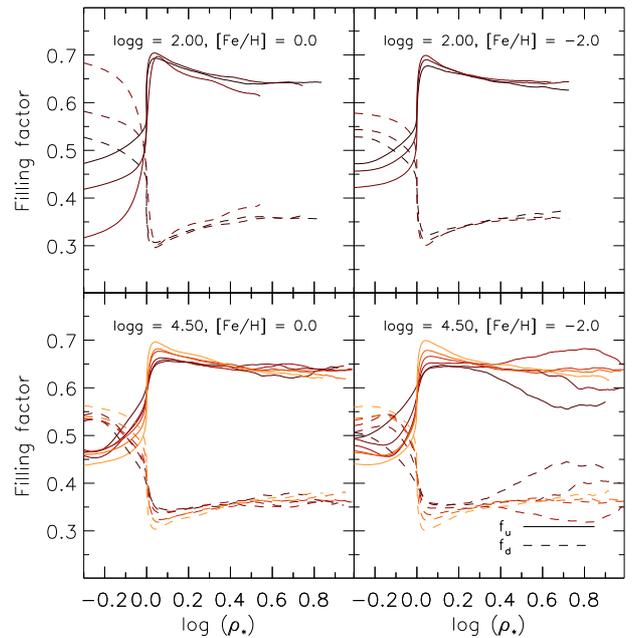}\caption{\label{fig:filling_factor}Filling factor of the up- and downflows
(solid and dashed lines) vs. normalized density $\rho_{*}$ for different
stellar parameters.}
\end{figure}
In Fig. \ref{fig:filling_factor}, we show the filling factors of
the up- and downflows. These are also very similar between the different
3D models. The upflows have a filling factor of $2/3$ and the downflows
$1/3$, which was already reported by \citet{1998ApJ...499..914S}
for the solar case. This means that on average the downflows are compressed
by a universal ratio of 2:1. This and the above findings are also
connected to the scaling of the granulation pattern, which we found
to scale with the pressure scale height in all \textsc{Stagger-grid}
models \citep[see][]{2014arXiv1405.7628M}, and suggests that surface
convection operates in the same way for all cool stars. Furthermore,
we note that the asymptotic character of the entropy jump is also
imprinted in the temperature and enthalpy, and becomes apparent when
decomposing them into the non-adiabatic and adiabatic parts.

\section{Fitting the normalized entropy jump\label{sec:Fitting-the-entropy}}

\begin{figure*}
\includegraphics[width=176mm]{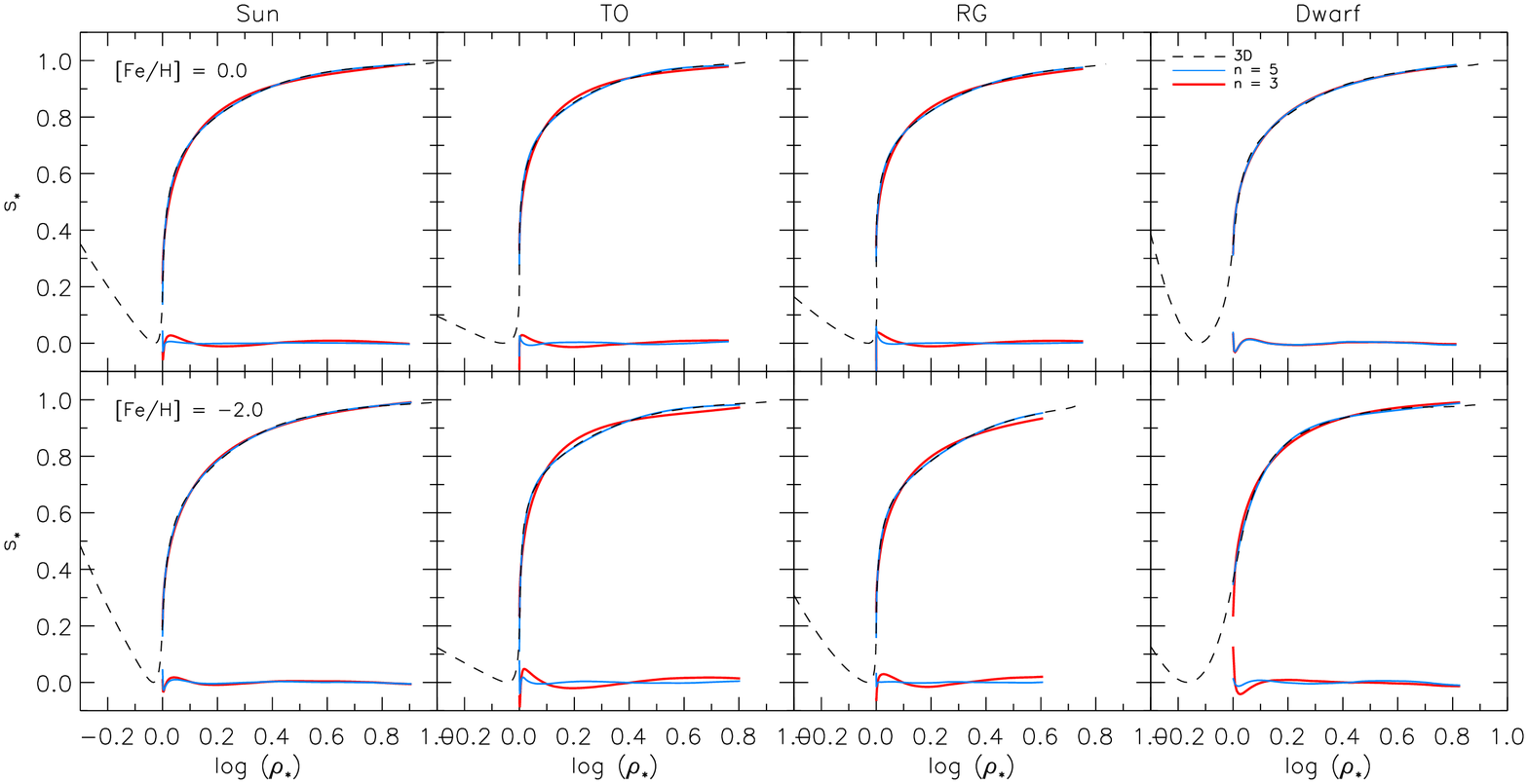}

\caption{\label{fig:Fitting-the-sun}Fittings of the normalized $\hav$ entropy
jump (dashed line) with the functions $\zeta_{3}$ and $\zeta_{5}$
(red and blue lines, respectively) for different models: Sun ($\teff/\logg$=5777/4.44),
turnoff star (6500/4.0), red giant (4000/2.0), and dwarf (4500/5.0)
for solar metallicity ($\feh=0$) and metal-poor ($\feh=-2.0$) models
(upper and lower panels). We also depict the difference between the
$\hav$ model and the fitted functions $\zeta_{3}$ and $\zeta_{5}$.}
\end{figure*}
\begin{figure}

\includegraphics[width=88mm]{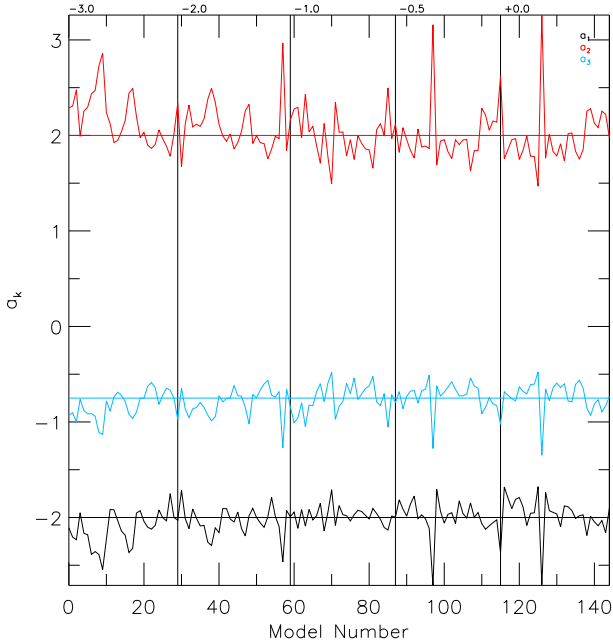}\caption{\label{fig:n3coeff}The different coefficients for $\zeta_{3}$ from
fits to different models. The different metallicities are indicated.}
\end{figure}
We find that the normalized entropy jump can be reasonably fit with
an asymptotic function, $\zeta_{n}$, of the functional form
\begin{eqnarray}
s & = & \zeta_{n}\left(x\right)\Delta s+s_{\mathrm{min}},\label{eq:entropy}
\end{eqnarray}
with $x=\log\rho_{*}$. A formulation for the $\zeta_{n}$ that matches
$s_{*}$ well is a generic polynomial:
\begin{eqnarray}
\zeta_{n} & = & 1+\sum_{k=1}^{n}a_{k}\left(1-x^{k/2}\right).
\end{eqnarray}
Initially, we applied an asymptotic function; however, the resulting
fits were insufficient, so that we found the polynomial to give a
better fit. In Fig. \ref{fig:Fitting-the-sun} we show the results
of the fitting for four different stellar models with two different
metallicities in comparison with the $\hav$ entropy stratification.
The differences, $\delta=s_{*}-\zeta_{n}$, are very small as shown.
In fact, the root mean square and the maximum of the differences averaged
over all models are $7.32\times10^{-3}$ and $9.45\times10^{-3}$
for the 3rd order function and $2.33\times10^{-3}$ and $3.61\times10^{-3}$
for the 5th order polynomial.

For the lower order, $n=3$, we find that the coefficients vary around
$a_{k}=-2,+2,-0.75$, as can clearly be seen in Fig. \ref{fig:n3coeff},
i.e. the function

\begin{eqnarray}
\zeta_{3}^{*} & = & 1-2\left(1-x^{1/2}\right)+2\left(1-x^{2/2}\right)-3/4\left(1-x^{3/2}\right)\label{eq:generic}
\end{eqnarray}
fits the normalized entropy jump on average. We find the smallest
residuals for $n=5$, as shown in Fig. \ref{fig:Fitting-the-sun}.
However, we cannot find any correlations between the fitting coefficients
and the stellar parameters. A better suited functional basis could
be found; in particular, at the bottom the function $\zeta_{n}$ does
not exactly exhibit an asymptotic behaviour by definition. Since an
asymptotic functional basis resulted in worse fits, we preferred to
use the presented generic polynomial.

\section{Depth dependent boundary for stellar structures\label{sec:Boundary-conditions}}

\begin{figure*}
\includegraphics[width=59mm]{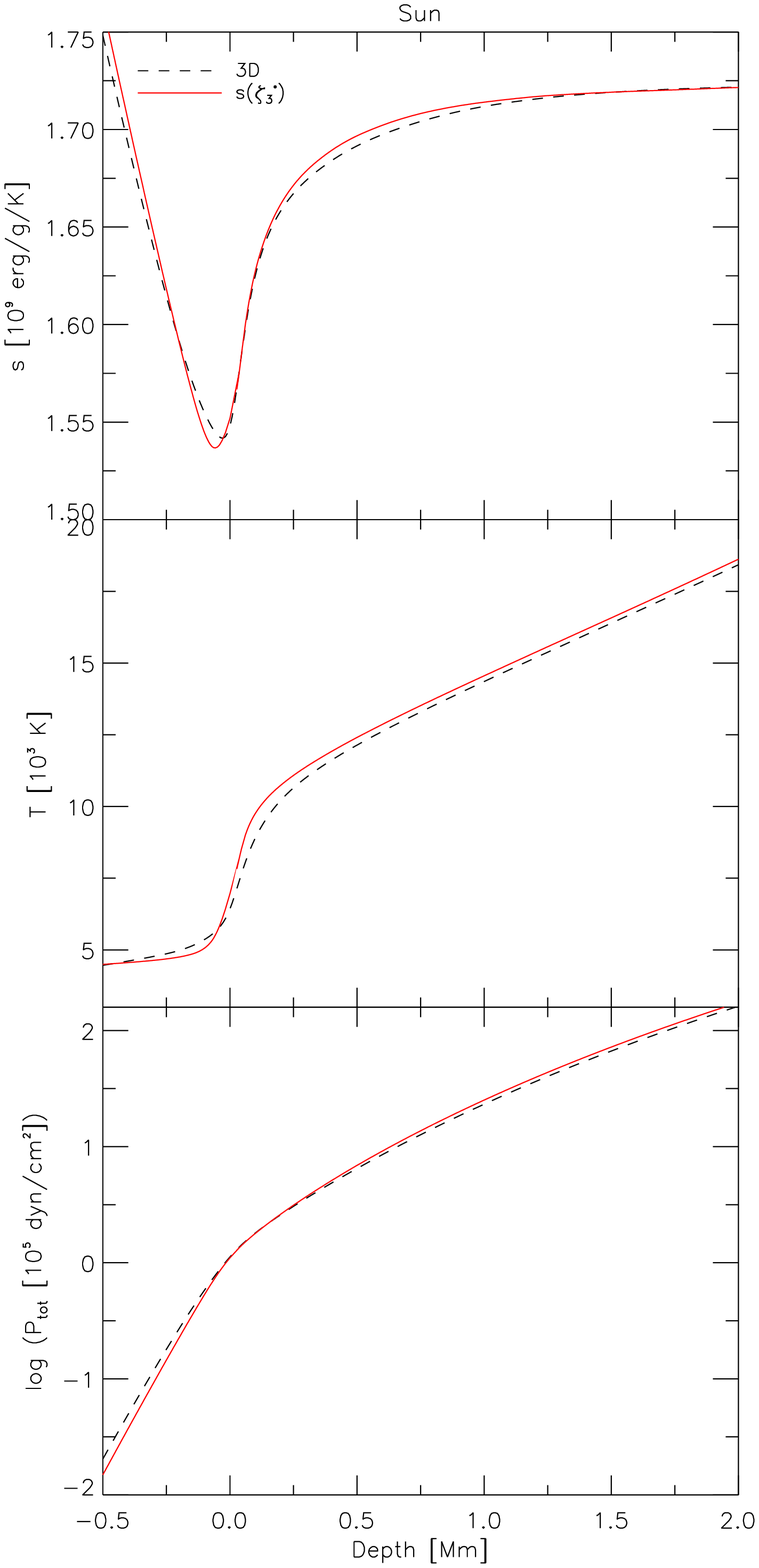}\includegraphics[width=59mm]{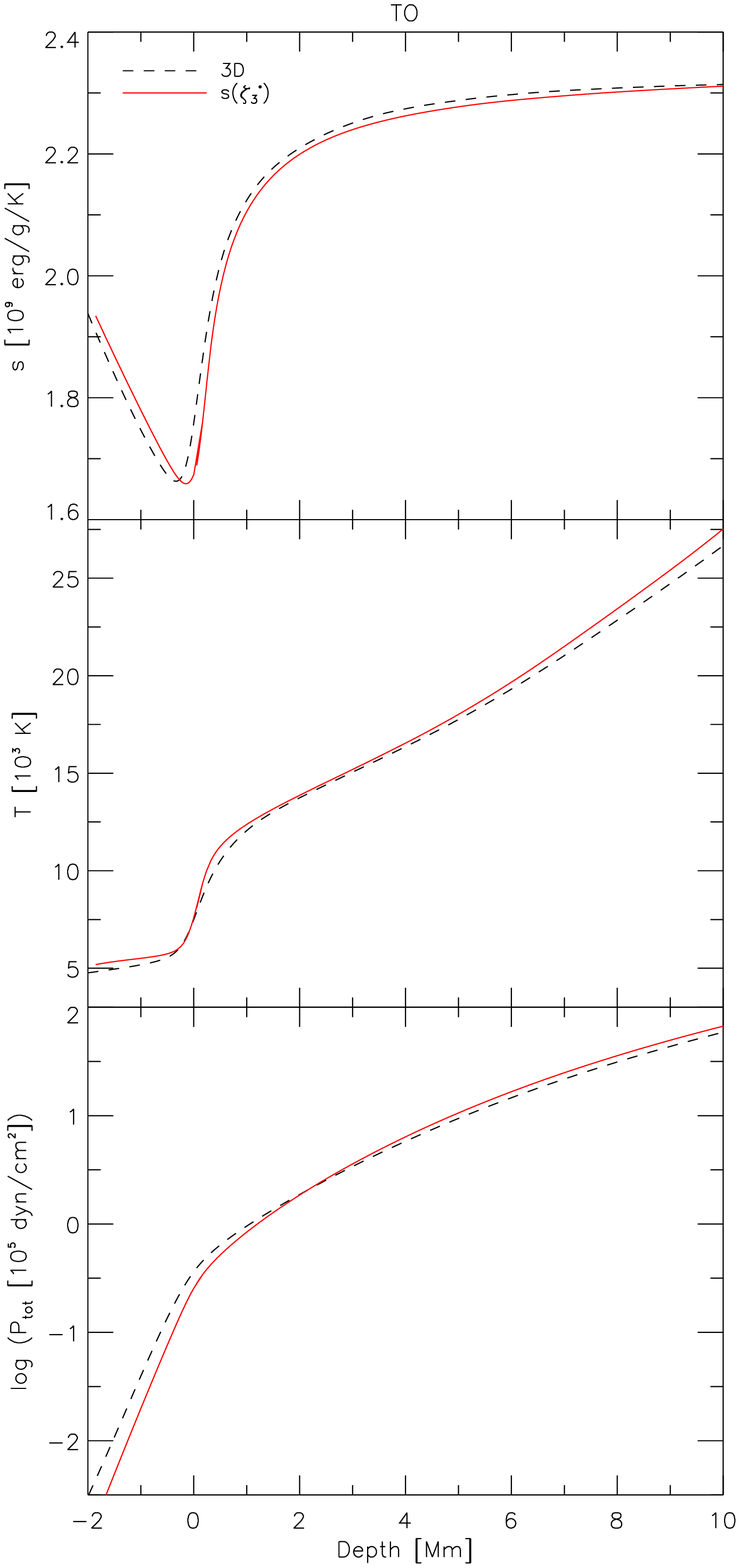}\includegraphics[width=59mm]{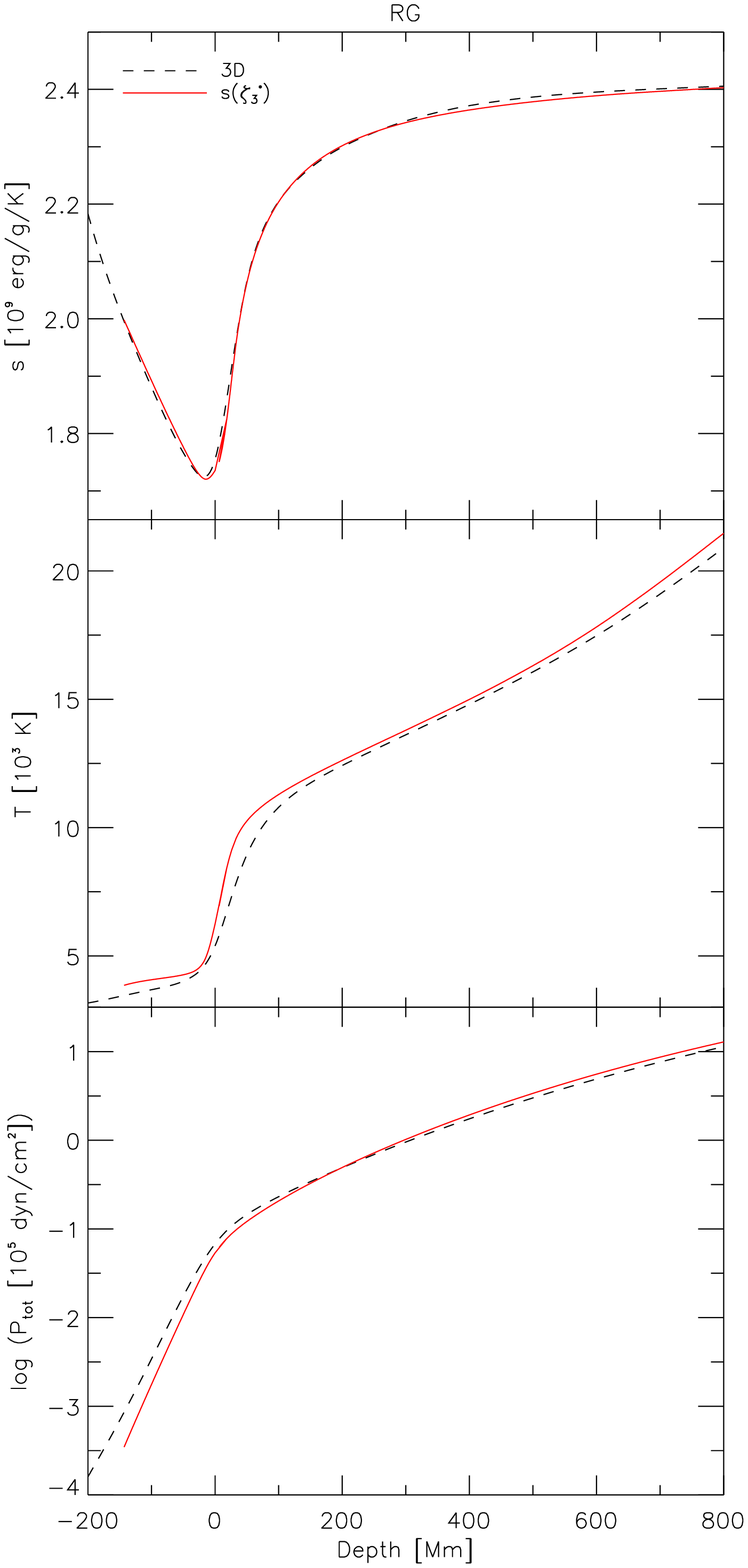}

\caption{\label{fig:reconst_sun}Reconstructed (\emph{red line}) stratifications
of the entropy, temperature and pressure (\emph{top, middle }and\emph{
bottom panels}, respectively) from the generic function $\zeta_{3}^{*}$
(Eq. \ref{eq:generic}) for the solar, red giant and turnoff models
(from left to right). We show also the $\hav$ model (\emph{dashed
line}) in comparison.}
\end{figure*}
In stellar structure calculations, the entropy can be conveniently
obtained by integrating 
\begin{eqnarray}
\frac{ds}{dz} & = & -\frac{c_{P}}{H_{P}}\left(\nabla-\nabla_{\mathrm{ad}}\right),
\end{eqnarray}
where $c_{P}$ is the specific heat at constant pressure, $H_{P}$
the pressure scale height, and $\nsad=\nabla-\nabla_{\mathrm{ad}}$
the superadiabatic gradient. Since the asymptotic entropy jump is
universal, we can construct the thermal stratification in the superadiabatic
layers using the generic entropy stratification scaled by the entropy
jump and minimum. In Fig. \ref{fig:reconst_sun}, we show a constructed
entropy stratification for the Sun using the generic function $\zeta_{3}^{*}$
given in Eq. \ref{eq:generic}. Compared with the $\hav$ model, the
differences between the two are remarkably small.

To construct such a stratification, we first consider the generic
relation between $s_{*}$ and $\rho_{*}$ from Eq. \ref{eq:generic}.
Then, for a given choice of stellar parameters $\teff$, $\logg$,
and $\feh$, we determine $\Delta s$ and $s_{\mathrm{min}}$, which
are provided in \citet{2013A&A...557A..26M} in the form of functional
fits. We note that the entropy minimum sets the outer boundary, while
the entropy jump determines the temperature gradient ($\nabla=\nsad+\nabla_{\mathrm{ad}}$).
With Eq. \ref{eq:entropy}, we can now scale the entropy stratification.
The density can be determined from Eq. \ref{eq:density_norm} and
the entropy peak, which is approximately located at the optical surface,
i.e. $ds|_{\mathrm{max}}\approx\tau_{\mathrm{Ross}}=2/3$. Therefore,
the adiabatic stratification with $s_{\mathrm{min}}$, where the entropy
is constant, can be helpful for an initial estimate of the value of
the density, at which $\rho\left(ds|_{\mathrm{max}}\right)$. From
the resulting $\rho$ and $s$, we compute the thermal pressure, $p_{\mathrm{th}}$,
from the equation of state. Then, the geometrical depth can be retrieved
from the hydrostatic equilibrium, $dp_{\mathrm{th}}/dz=\rho g$. We
note that the surface gravity is responsible for the scaling of the
geometrical depth. The turbulent pressure (and velocity) has to be
included in the hydrostatic equilibrium because it will elevate atmospheric
structure (i.e. $p_{\mathrm{tot}}=p_{\mathrm{th}}+\rho u_{z}^{2}$).
However, for simplicity we neglect the latter. The location of the
optical surface can be retrieved from the optical depth, $d\tau/dz=\rho\kappa$.
Finally, with the constructed entropy and density stratification one
can compute other thermodynamic quantities, such as temperature or
internal energy, from the equation of state. We note that in the different
constructed models shown in Fig. \ref{fig:reconst_sun}, we appended
an isothermal atmosphere above the surface.

This procedure can be used to construct depth dependent thermal stratifications
to be used as outer boundary conditions for stellar structure computations.
The physical complexity of the radiative transfer equation is then
hidden in the dependence of $\Delta s$ and $s_{\mathrm{min}}$ on
stellar parameters. This approach would reduce the deficiencies of
MLT and the $\ttau$ relations since the mixing length is encoded
in $\Delta s$ and the outer boundary in $s_{\mathrm{min}}$. Such
a simplified prescription of the superadiabatic layers could possibly
improve the $p$-mode frequency disagreement \citep{1988Natur.336..634C,1999A&A...351..689R}.
We note that our aim in the presented study is to illustrate the potential
application of our findings for stellar structures. The method summarized
above clearly needs further development for improved results.

\section{Conclusions\label{sec:Conclusions}}

The presented scaling relation for the entropy jump indicates that
radiative losses in cool stars takes place in a fairly similar fashion.
We found that the normalized entropy, as a function of the normalized
density, can be fitted very well with a polynomial function. A robust
and generic, yet simple description of the normalized entropy stratification
can be achieved with a three-parameter function. This could be helpful
for stellar structure computations by providing a simple description
of the otherwise non-trivial superadiabatic region in cool stars.
In particular, as corrections for the near-surface effects in asteroseismology
this would constitute a large step forward in stellar structure modelling.
We have shown that from the generic normalized entropy, we can construct
stratifications by scaling the normalized entropy with the entropy
jump and minimum, as outlined above. However, the method can benefit
from some refinement in the future, for example by including the turbulent
pressure in the hydrostatic equilibrium equation. Superadiabatic convection
seems to take place in a similar manner, despite large differences
in the physical conditions. The reason behind this has yet to be explored.
\begin{acknowledgements}
This work was supported by a research grant (VKR023406) from VILLUM
FONDEN.
\end{acknowledgements}

\bibliographystyle{aa}
\bibliography{papers}

\end{document}